\documentclass   [a4paper,12pt]{article}
\usepackage{graphicx}
\begin{document}
\begin{center}
{\large\bf PHENOMENOLOGICAL AND ONTOLOGICAL MODELS \\
                           IN NATURAL SCIENCE} \\[2mm]

{\small Milo\v{s} V. Lokaj\'{\i}\v{c}ek}, 
{\small Institute of Physics, AVCR,} 
{\small 18221 Prague 8, Czech Republic}\\
{\small e-mail: lokaj@fzu.cz}  
\end{center}

\begin{abstract}
The observation of the nature and world represents the main source
of human knowledge on the basis of our reason. At the present it
is also the use of precise measurement approaches, which may
contribute significantly to the knowledge of the world but cannot
substitute fully the knowledge of the whole reality obtained also
with the help of our senses. It is not possible to omit the
ontological nature of matter world. However, any metaphysical
consideration was abandoned when mainly under the influence of
positivistic philosophy phenomenological models started to be
strongly preferred and any intuitive approach based on human
senses has been refused. Their success in application region has
seemed to provide decisive support for such preference. However,
it is limited practically to the cases when only interpolation
between measured data is involved. When the extrapolation is
required the ontological models are much more reliable and
practically indispensable in realistic approach.
\end{abstract}

\section{Introduction}
Human knowledge is based on the ability of human reason. One
starts from the pieces of knowledge obtained by observation of the
world (including human existence). All scientific knowledge is
based fundamentally on such an approach. The human reason forms
some generalized statements (or hypotheses) from the individual
pieces with the help of logical induction (event. of intuition).
All possible consequences are then to be derived from all these
propositions with the help of logical deduction. The statements that
do not lead to any mutual contradictions and to any contradictions
with world observation may be denoted as plausible. In the case of
a contradiction the given hypothesis (or a given set of
hypotheses) must be denoted as falsified. It  is not possible to
start from it in extending our knowledge as far as it is not
modified in corresponding way. However, the propositions that have not been
falsified may be denoted only as plausible. It is not possible to
speak about the verification of scientific theories; see
\cite{lok1}.

The main role is played in this approach by natural science,
especially by physical, chemical, and also biological research.
Such research might be hardly possible without the help of
mathematical models. The models used at the present should be
divided into two categories: phenomenological or ontological
(denoted in biology usually as mechanistic). We will discuss in
the following the questions concerning the possibility of using
different models in extrapolation predictions and of their
contributing to better understanding of natural phenomena.

However, in Sec. 2 we will mention first shortly the possibility
of human knowledge and the problem of opinion plurality and in
Sec. 3 the successive development of scientific knowledge. The
evolution of thinking with the coming of the new age will be
mentioned in Sec. 4. The mathematical modelling in natural
sciences will be handled in Sec. 5. The necessity of metaphysics
will be stressed in Sec. 6. And the problem of intuitive knowledge
in the region of microscopic world will be analyzed in Sec. 7.
Phenomenological characteristics remaining still necessarily in
contemporary basic natural science will be mentioned in Sec. 8.

\section{Human knowledge and tolerance problem}

It has been introduced already in the preceding section that our
knowledge is based on falsification approach. Any positive
statement (proposition) may be shown or as falsified or as plausible. It may be
never denoted as verified. Consequently, two (or more) mutually
contradicting statements may be denoted as plausible if any of
them has not been falsified in tests performed in
connection with an additional plausible statement set.

That may be denoted as the source of fully legitimate plurality of
different opinions (or also theories). The legitimate and fully
justified tolerance must be exhibited towards such statements. It
means that also two alternative (different) theories should be
taken as plausible if they have not been falsified in separate
tests. However, that is practically excluded in the contemporary scientific
approaches on the basis of falsifiability principle, even if on
the other side, the equivalent tolerance is often required in
other regions (e.g., in the region of metaphysics) also for
statements that have been already falsified in the past. Such
tolerance must be denoted as destructive, as it tends necessarily
to an untrue picture of the world or human being.

From the fact that the way to knowledge consists in the
falsification approach it might seem that practically
nothing may be known with certainty. That holds actually for any
non-falsified positive statement; it can never be said that it
holds actually and will hold also in the future. However, the
certain knowledge may exist; it consists of the whole set of all
falsifying statements. The certainty may be obtained always about
what does not hold; therefore, about what is not true.

\section{Scientific knowledge and its evolution}
The people were forced to learn to know the nature from the very
beginning of their existence to save or to lighten their lives.
However, the actual beginning of systematic (scientific) knowledge
may be put into the old Greece approximately in the fifth century
b. C. The matter structure belonged always to basic
considerations. Demokritos (cca 470-360 b. C.) formulated the
hypothesis about the existence of atoms (i.e., of furthermore
indivisible small bodies) that fill fully their internal space
which is assumed to be otherwise empty. From Leukippos he took over
also the law of causality: Any thing does not arise without a
cause, but everything arises from some reason and necessity. Such
intuitive ideas and standpoints did not appear practically in
other parts of the world.

The given intuitive approach was conserved also in the
metaphysical (ontological) approach of Aristotle  (384-322 b. C.),
who put also the grounds to whole natural science on the basis of
world observation, even if his own statements about the world
(starting from the then observations) were later falsified.
Aristotle did not limit himself, of course, only to natural
science based on direct world observation. Indivisible part of his
contribution was the metaphysical approach, when he formulated
also different conclusions following from the ontological nature
of real matter objects. He formulated also the rules of two-value
logic on such a basis and developed further the causality principle.

However, the approaches of Aristotle were not known in Europe
practically for the whole thousand of years a. C. Islam brought
them to Europe only in the beginning of the second millennium.
They were taken over by Albert Magnus (1193-1280) and also by
Thomas Aquinas (1225-74), who then continued in developing further
all previous philosophical aspects. Thomas was developing, of
course, also theological philosophy that tended from the Christian
revelation towards the matter world and had, therefore, a source
differing from that of metaphysics. It is then possible to say
that Thomas was distinguishing in principle always between these
two different philosophical regions. And the natural science was
developing at that time always in full harmony with metaphysical
consideration, practically till the beginning of the new age.

It is possible to say that successive changes occurred when the
followers of Thomas Aq. were not able to continue in keeping the
metaphysical consideration in agreement with nature observations
and took some metaphysical statements as dogmas. And also some
statements of natural sciences were not corrected in agreement
with contemporary observations and remained under the influence of
different earlier convictions; e. g., F. Bacon (1561-1626) was
speaking about different idola (mistakes). At that time the
importance of falsification approach was, of course, not yet known
and realized.

\section{Evolution of scientific thinking in the new age}
A series of different ideas appeared gradually in the new age that
were accepted as new starting points for formulating corresponding
scientific hypotheses. It is possible to say that mainly three
newly formulated concepts influenced the development of the modern
physics and whole natural science. It was the overestimation of
the human reason by R. Descartes (1596-1650), the refusal of
causality and the overestimation of chance in natural evolution by
D. Hume (1711-76) and the positivist philosophy of A. Comte
(1798-1857) who limited all considerations to mere measured data
and refused any metaphysical thinking. It was probably also the
discovery of molecular and atom structures in the 18th century
that contributed at least partially to these trends as the
probability distribution started to play an important role in the
interpretation of physical phenomena.

It is then possible to say that the origin of two main physical
theories of the 20th century (i.e., Copenhagen quantum mechanics
and special relativity theory) was influenced fundamentally by E.
Mach (1838-1916), who refused any metaphysics in physical
considerations \cite{holton}. Thus the purely phenomenological
mathematical models were strongly preferred and the modern physics
was built up practically on them.

As to the quantum mechanics it is well known that the contemporary
model leads to logical quantum paradoxes and that until now the
sharp gap between the microphysics and macrophysics has not been
removed, either. We have analyzed the basis of quantum mechanics
and its reasoning to a much greater detail in the last time; see
[1,3] and papers quoted there. And we have shown that the
Schr\"{o}dinger equation itself (without any additional
assumptions) must be preferred to the Copenhagen quantum mechanics
as well as to mere classical physics, even if there is not any
difference in the description of stable objects with the help of
Schr\"{o}dinger or classical equations. Thus, the so-called
hidden-variable theory represents the best description of physical
reality (see also Sec. 5). And we may conclude that also in the
microscopic world the corresponding ontological model is to
substitute the purely phenomenological Copenhagen mathematical
model. All earlier quantum paradoxes disappear and the microscopic
physics may be interpreted in intuitive ontological way as the
macroscopic one. Also the interpretation gap between microscopic
and macroscopic physical regions disappears \cite{lok2}.

The preference of ontological models in natural science seems to
be strongly supported in such a case, which will be discussed to a
greater detail in the next section.

\section{Mathematical modeling of natural processes}
It is possible to say that any research concerning the matter
world (including human society) cannot manage without using
mathematical models, being partly very simple and partly very
complex. They were introduced into the life by G. Galileo
(1564-1642) and I. Newton (1643-1727). Their models may be denoted
as ontological. Sometimes it is spoken about such models as about
mechanistic ones, to stress that they should represent matter
mechanism hidden behind, at the difference to purely
phenomenological representation of measured values.

The contemporary physical theories are based, of course,
practically fully on phenomenological models, without devoting any
interest to deeper matter (ontological) mechanisms. One obtains
mostly quite non-intuitive description of corresponding phenomena
in such cases. The quantum paradoxes (wave-particle duality,
non-localization of microscopic particles, tunnel phenomenon, and
similarly) have been derived with the help of the phenomenological
Copenhagen quantum-mechanical model. Such phenomena have been
denoted oft as the reality of the microscopic world, differing
significantly from the macroscopic one.

Any objections against phenomenological quantum-mechanical models
and against corresponding consequences have been refused by
arguing usually that these models have been very successful in
passing to technological applications. It holds, however, to the
extent when the interpolation in the region of measured values is
being performed. Different unsuccessful cases may be shown if the
extrapolation outside the region of measured values has been done.
Such failure has not been exhibited as a rule when ontological
models have been made use of.

The abandoning and refusing of ontological approaches must be
denoted as a mistake also from another reason. It would not be
possible to speak about scientific thinking or approach if the
already mentioned two-value logic were not involved. And Aristotle
proposed just this kind of logic starting in principle from the
ontological basis of the matter world.

Some efforts to introduce more-value logics appeared in the 20th
century. However, the attempts to apply them to natural phenomena
in connection with quantum mechanics were unsuccessful. And there
is not any reason for it, either, since we have shown that the
quantum theory must be based on the mere Schr\"{o}dinger equation
(as introduced already in Sec. 4), which provides the results for
stable microscopic objects to be practically equivalent to
classical physics [3]. It is the theory with hidden parameters
supported all the time by A. Einstein that must be applied to if
the microscopic phenomena are to be described in agreement with
reality [1,3].

\section{Physics and metaphysics}
As there is not any reason for a more-value logic the way is again
fully open for applying ontological considerations in extending
our knowledge. Especially, when commonly used standard logical
rules represent practically indivisible part of any scientific
approach.

It is also the intuitive insight that is to come back into the
natural science. It may be hardly possible to interpret further
the real world by manipulating mathematical artifacts that are
usually linked up with phenomenological models. There is not any
doubt that the actual knowledge of the world should be connected with
ontological models. One is entitled to expect that a similar
synthesis of scientific and philosophical (metaphysical) thinking
may occur again as it was in the time of Aristotle and also of the
middle age.

However, the other philosophical region (i.e., theological
philosophy) introduced in the middle age need not be influenced
very much in such a case as it has a quite different starting
point than natural science. It starts from the Judaic-Christian
revelation that is contained in the books of Old and New
Testament. The theological philosophy combines, of course, its
conclusions also with the pieces of knowledge obtained on the
basis of world observation, which means that we are entitled to
apply the standard falsification approach to corresponding
theologically oriented statements as far as they concern the world
and human being. It means that also in this region some plurality
of mutually contradicting opinions may exist when none of them was
falsified separately. It is possible to lead a dialogue between
corresponding statements in looking for a true answer; a kind of
dialectic synthesis may be formulated when the whole known
non-falsified "truth" is taken into account and falsification does
not occur. On the other side some new hypotheses concerning the
matter world might be initiated by corresponding considerations.

\section{Microscopic world and sense knowledge}
One region of the physics of the 20th century was devoted mainly
to the description of the microscopic world. It was based
practically on phenomenological models and the derived picture of
the microworld was very different from that might be obtained by
extrapolating the experience gained with the help of our senses.
The ontological models have commemorated the situation of the
intuitive approach started by Demokritos (as mentioned already in
Sec. 3) or developed further mainly by Aristotle and Thomas Aq.
The given picture was supported in principle also by discoveries
in the 17th and 18th centuries when the actual existence of atoms
and molecules was discovered.

However, at that time different trends started to appear when
instead of ontological models the phenomenological ones were made
use of. Important example may be represented by the discovery of
entropy increase when this phenomenological characteristic was
denoted practically as a natural law or as the cause of
corresponding behavior of many-bodies systems even if it should be
interpreted as the consequence of other rules holding for
interactions of individual bodies under given conditions. The
situation did not change even if the entropy increase was
substituted by the probability increase of atom (or molecule)
distribution. The given evolution continued when the causality
principle was refused and the chance (and probability) was
declared as the basic characteristic of natural evolution, as Hume
proposed it.

Phenomenological models represented always important lost of
intuitive insight as to the matter structure. N. Bohr contributed
then further to such trends when in 1913 he formulated two
postulates concerning the atom structure \cite{bohr}; two
phenomenological characteristics were denoted in principle as
physical laws being fully responsible for corresponding physical
behavior. To the given evolution also L. de Broglie \cite{brogl}
contributed when he attributed a pilot wave to any mass particle,
for which any ontological reasoning did not exist. However, N.
Bohr brought a decisive contribution to the support of
phenomenological models, when he introduced the Copenhagen quantum
mechanics \cite{bohr2}. It was based on time-dependent
Schr\"{o}dinger equation [7], to which, however, some further
additional important (in principle only mathematically formulated)
assumptions were added. The given theory has led then to some
paradoxical properties that have been denoted as
actual properties of microworld.

It has been often stated that many experiments have entitled us to
take the Copenhagen quantum mechanics as actually valid. However,
it has been almost always only the mere Schr\"{o}dinger equation
that has been tested in corresponding experiments; none of the
mentioned additional Copenhagen assumptions was actually taken
into account in such tests. And it has been already shown that the
basic solutions (i.e., containing only one Hamiltonian
eigenfunction) of the time-dependent Schr\"{o}dinger equation
have led to the results that are fully equivalent to those of
classical physics \cite{lok2}. It does not hold for superposition
solutions (involving at least two Hamiltonian eigenfunctions),
which makes the Schr\"{o}dinger equation more general than the
equations of classical physics. That is in full agreement with the
conclusions of Hoyer \cite{hoyer} and Ioanidou \cite{ioan} who
showed that the Schr\"{o}dinger equation may be derived if the
classical physics is combined with a statistical distribution of
some basic physical parameters. There are then some experimental
facts (existence of discrete states) indicating that the
Schr\"{o}dinger equation should be preferred to other model
alternatives in the region of microscopic world \cite{lok2}.

One may say that the EPR experiment measuring two-photon
coincidence was the only one where the mentioned additional
Copenhagen assumptions were actually discussed and practically tested.
However, the violation of Bell's inequalities found experimentally
by Aspect et al. \cite{asp} has been interpreted in such a way
that two untrue statements have been involved; see [1] and papers
quoted there. In fact only a simplified hidden-variable theory may
be excluded together with classical alternative by the given
experimental data; the corresponding violation has not concerned
the general hidden-variable alternative (or the general 
Schr\"{o}dinger equation). It means that nothing prevents us from
returning to intuitive ontological picture of matter structures in
the region of microscopic phenomena.

\section{Ontological basis and some remaining \\ phenomenological characteristics }
Nevertheless, it is necessary to introduce that the physics is
still fully dependent on one quite phenomenological quantity. It
is the distant force action. One may easily imagine and interpret
ontologically the action of a force (repulsive as well as
attractive) at very small distances (objects being in contact).
The force action at a distance (electromagnetic and gravitation
forces) may be understood until now on phenomenological level
only, even if it may be practically clearly defined on the basis
of corresponding consequences. There is a quite open (till now
unanswered) question how it might be possible to interpret it
ontologically as other physical characteristics. The
phenomenological description may be, however, made use of, too,
when the ontological mechanism of some part of reality is very
complex and for the analysis of the whole system a
phenomenological relation is sufficient.

On the other side it is hardly possible to accept an
interpretation, even if it is ontologically based, if some
violation of logical principles is involved. Such a problem
concerns the strong interactions (between hadron particles) that
may be regarded in principle always as contact interactions. At
the present the hadrons are assumed to consist of smaller quarks
and all mutual interactions between them are interpreted on the
basis of interactions between individual quarks. It follows from
experimental data and is practically certain that hadrons cannot
be regarded as elementary objects; they must consist of some
smaller parts, having been denoted commonly as partons. However,
one should ask whether these partons may be identified with
contemporary quarks as it is generally assumed.

The quarks are assumed now to have relatively low rest masses and,
therefore, they should be loosed normally as free objects in
hadron collisions at sufficiently high energies. However, only
many other secondary hadrons (or their resonances) have been
always found in such collisions. As there has not been any direct
evidence for the quark existence one special property has been
attributed to them, which may be described mathematically but it
may be hardly understood ontologically and intuitively. When
brought to high velocities (as parts of an accelerated hadron) and
to interactions with other quarks (present in a different
colliding hadron) the pair of mutually interacting quarks has been
assumed to be unable to withdraw oneself from another one. They
have been expected to remain mutually bound and only the process
of their hadronization (i.e., transmutation to hadrons) has been
assumed to occur. And one must assume further that even many tens
of secondary hadrons may be formed from one couple of colliding
quarks. There is not any description, either, what occurs with
other quarks that do not collide directly.

No explanation (or concrete description) has been given until now
for such unusual phenomenological quark behavior, even if already
a very long time has passed from the time when the given
assumption was formulated. And so, even if at the first sight the
existence of quarks might represent quite intuitive approach one
is forced to ask whether their mysterious properties (or rather
the mathematical artifacts) do not disqualify them actually from
further considerations. The usual argument that any other
explanation does not exist until now should be denoted as
contra-productive (having been untrue at the same time).

There has been the question put before several decades of years
and not answered until now: whether the secondary hadrons and
hadron resonances (appearing in hadron collisions) are not
preformed in individual hadrons before being loosed in a
collision. The mentioned partons might be then identified with
these resonances. However, in the past they were identified in
principle with individual quarks representing basic states in
SU(3) theory.

The partons (similarly as all hadrons) might then consist of other
smaller objects (i.e., of some quasi-quarks). These quasi-quarks
should be then very heavy being mutually strongly bound at very
small distances. Different changeable parton structures might be
then formed inside individual hadrons, while individual partons
(i.e., contemporary resonances) might be get free in hadron
collisions by different stripping reactions, which would be
followed by the decay of highly excited objects. Very heavy
resonances (corresponding to partons) might be then formed also in
direct mutual collisions of partons existing in colliding hadrons
when also some quasi-quark pairs should be created (or again
annihilated). Jets consisting of many secondary hadrons might be
then formed by successive decays of highly excited resonances.
Thus, the purely phenomenological quantum chromodynamics might be
substituted by a model more acceptable from purely logical as well
as ontological views.

What remains open is the nature of the corresponding quasi-quarks
and of their mutual interactions. The forces between them must
change strongly with their mutual distances and act mostly in a
collective manner. And just this property should be described in a
quite new way. It would be necessary to describe (or explain) also
the changeable parton structure inside individual hadrons during
their free existence between mutual collisions and interactions on
the basis of such quasi-quark property.

\section{Conclusion}
There is not any doubt that the modern physics would not be so
successful without making use of mathematical models. However, one
must ask whether the contemporarily used phenomenological models
may really contribute to the true knowledge of the world.
Especially, if they are in disagreement to the knowledge gained
with the help of human senses.

It is evident that our sensual ability leads always to fully
ontological view, which is in agreement with the fact that no
science would be possible without using the two-value logic based
on realistic ontological view. It has been introduced in Sec. 7
that especially the phenomenological model of Copenhagen quantum
mechanics has not been based on realistic grounds. The way has
been opened to substitute it by a fully ontological
hidden-variable model \cite{lok2}. There is not more any actual
reason for making use of purely phenomenological models at the
present time, either.

\end{document}